# Giant Biquadratic Exchange in 2D Magnets and its Role in Stabilizing Ferromagnetism of NiCl$_2$ Monolayer

J. Y. Ni$^{1,2*}$, X. Y. Li$^{1,2*}$, D. Amoroso$^3$, X. He$^4$, J.S. Feng$^{1,5\dagger}$, E. J. Kan$^6$, S. Picozzi$^3$, and H. J. Xiang$^{1,2,7\dagger}$

$^1$*Key Laboratory of Computational Physical Sciences (Ministry of Education), State Key Laboratory of Surface Physics, and Department of Physics, Fudan University, Shanghai 200433, P. R. China*

$^2$*Shanghai Qizhi Institution, Shanghai 200232, P. R. China*

$^3$*Consiglio Nazionale delle Ricerche CNR-SPIN Via dei Vestini 31, Chieti 66100, Italy*

$^4$*Catalan Institute of Nanoscience and Nanotechnology (ICN2), CSIC, BIST, Campus UAB, Bellaterra, Barcelona, 08193, Spain*

$^5$*School of Physics and Materials Engineering, Hefei Normal University, Hefei 230601, P. R. China*

$^6$*Department of Applied Physics and Institution of Energy and Microstructure, Nanjing University of Science and Technology, Nanjing, Jiangsu 210094, P. R. China*

$^7$*Collaborative Innovation Center of Advanced Microstructures, Nanjing 210093, P. R. China*

*These authors contributed equally to this work.

†email: fjs@hfnu.edu.cn

†email: hxiang@fudan.edu.cn

**Abstract-**Two-dimensional (2D) van der Waals (vdW) magnets provide an ideal platform for exploring, on the fundamental side, new microscopic mechanisms and for developing, on the technological side, ultra-compact spintronic applications. So far, bilinear spin Hamiltonians have been commonly adopted to investigate the magnetic properties of 2D magnets, neglecting higher order magnetic interactions. However, we here provide quantitative evidence of giant biquadratic exchange interactions in monolayer NiX$_2$ (X=Cl, Br and I), by combining first-principles calculations and the newly developed machine learning method for constructing Hamiltonian. Interestingly, we show that the ferromagnetic ground state within NiCl$_2$ single layers cannot be explained by means of bilinear Heisenberg Hamiltonian; rather, the nearest-neighbor biquadratic interaction is found to be crucial. Furthermore, using a three-orbitals Hubbard model, we propose that the giant biquadratic exchange interaction originates from large hopping between unoccupied and occupied orbitals on neighboring magnetic ions. On a general framework, our work suggests biquadratic exchange interactions to be important in 2D magnets with edge-shared octahedra.

***Introduction***-Reduced dimensionality and interlayer coupling in van der Waals (vdW) materials give rise to intriguing electronic, optical and other quantum properties [1-5]. In this general context, two-dimensional (2D) vdW magnets have recently emerged as an exciting class of materials with appealing perspectives for low-power miniaturized spintronic devices. For example, atomically-thin layers of $CrI_3$ and $CrGeTe_3$ were reported to be intrinsic ferromagnetic (FM) insulators[6,7], with significant Kitaev interactions (one type of anisotropic symmetric exchange)[8,9]: not only are the latter relevant to the magnetic anisotropy, but also likely result in the exotic quantum spin liquid state under strain[10]. VdW magnets are also an ideal playground for investigating topological states such as antiferromagnetic (AFM) topological insulators[11-13]. In order to elucidate the underlying physical mechanisms and to predict the properties of 2D magnets (e.g., $CrI_3$, $CrGeTe_3$, $VSe_2$, $Fe_3GeTe_2$, $VI_3$, $MnBi_2Te_4$)[6,7,11,14-17], bilinear spin Hamiltonians (including second-order single-ion anisotropy, Heisenberg symmetric exchange, Dzyaloshinskii–Moriya anti-symmetric exchange, and anisotropic symmetric exchange[18]) were commonly employed and found to accurately describe many of the 2D magnets[8,10,19-22]. However, an appealing question of general interest is whether higher-order spin interactions play a decisive role on the magnetic properties of some layered magnetic materials.

Among 2D magnets, we here focus on transitional metal dihalides $NiX_2$ (X = Cl, Br and I), where each single-layer of $NiX_2$ - separated from neighboring layers by vdW gaps-contains triangular nets of Ni-cations in edge-shared octahedral coordination. Due to the octahedral crystal field that splits Ni $3d^8$ orbitals into filled $t_{2g}$ states and half-filled $e_g$ states (*i.e.* only the $e_g$ majority spin channel is filled), $NiX_2$ are semiconducting systems with magnetic moment of ~$2\mu_B$ per Ni. Their magnetic properties are particularly intriguing. Neutron powder experiments on bulk systems reported helimagnetic layers of Ni-spins in $NiBr_2$ and $NiI_2$[17], whereas spontaneous electric polarization induced by the helical arrangement was later reported in Refs.[23,24], thus proving $NiBr_2$ and $NiI_2$ to be typical type-II multiferroics[25-28]. Conversely, for bulk $NiCl_2$ ferromagnetic layers were reported, showing a weak interlayer antiferromagnetic coupling; the application of a small magnetic field in bulk $NiCl_2$ can induce a transition from interlayer antiferromagnetism to ferromagnetism[29]. Recently, it was also proposed that 2D $NiCl_2$ in the FM state is a half-excitonic insulator [30] which may host a new type of Bose-Einstein condensation. So far, however, it has not been investigated why NiCl$_2$ displays ferromagnetic order, at variance with non-collinear magnetic ground states in NiBr$_2$ and NiI$_2$, despite all the three Ni-halides sharing a similar structure[17,31].

In this Letter, we investigate the peculiar magnetic properties of a single-layer of NiCl$_2$ and discuss them in relation to corresponding NiBr$_2$ and NiI$_2$ monolayers, relying on a spin Hamiltonian constructed by means of our recently developed machine learning method based on first-principles calculations[32]. We found that the usually overlooked biquadratic exchange is the key interaction for the stabilization of the FM collinear order within NiCl$_2$ layers; in particular, the strong and negative nearest-neighbor (NN) biquadratic coupling overcomes the weak magnetic frustration, induced by competing FM first-neighbor and AFM third-neighbor exchange interactions. Moreover, to understand the microscopic origin of such a giant biquadratic exchange interaction, we resort to a three-orbital Hubbard model, replacing the electron configuration with a hole picture. We found that the hopping between the occupied orbital of a site and the empty orbital of the neighboring site is essential to the giant biquadratic interaction, along with the coordination geometry: the 90° Ni-X-Ni geometry shows a much larger biquadratic exchange with respect to the 180° configuration. Finally, besides $d^8$ (e.g., Ni$^{2+}$ ions) systems, we predict a giant biquadratic exchange interaction also in $d^3$ systems, such as prototypical CrI$_3$ and CrGeTe$_3$ with Cr$^{3+}$ ions.

*Results*- Usually inter-layer magnetic interactions across the VdW gaps in vdW magnets are weak [29,31]. Indeed, our DFT calculations (not reported here) show that the inter-layer interaction has negligible effect on the intra-layer exchange couplings. In the aim of identifying the underlying mechanisms behind the observed FM order within NiCl$_2$ single-layers, we therefore investigate the magnetic properties of a free-standing monolayer. When computing exchange interaction parameters by employing DFT and including spin-orbit coupling (SOC) (see **Table S2-3** and the "Computational Details" Section in Supplementary materials(SM)[33]), we found out isotropic behavior for NiCl$_2$ (in agreement with previous experimental [29] and theoretical works[34]), at variance with NiBr$_2$ and NiI$_2$, where SOC-induced anisotropies are relevant. In order to test the effects of spin-lattice as well as SOC for NiX$_2$ monolayers, we additionally calculated via DFT the relative energies for various common commensurate magnetic structures [FM, NAFM, SAFM, DAFM and ZAFM, see **FIG.1** and **FIG.S1**] using three different procedures [35]. As shown in the **FIG.1** and **FIG.S1**, the three procedures give similar energetics, suggesting that spin-lattice coupling and SOC can be safely neglected for our purposes. Indeed, as the focus here is the collinear spin ordering in $NiCl_2$ and in order to allow a consistent comparison of the estimated magnetic interactions of $NiCl_2$, $NiBr_2$ and $NiI_2$, we'll consider from now on only isotropic contributions to the spin Hamiltonian.

In order to identify stable zero-temperature incommensurate magnetic configurations in monolayers $NiX_2$, we calculated the DFT total energy (neglecting SOC) as a function of the magnetic propagation vector ***k***, by employing the generalized Bloch theorem (gBT)[36] as implemented in the VASP package[37,38]. The dependence of the total energy $E(\boldsymbol{k})$ upon $(k_x, k_y)$ for monolayer $NiCl_2$, $NiBr_2$ and $NiI_2$ are presented in **FIG.2**(a)-(b) and **FIG.S7**. For monolayer $NiCl_2$, the energy minimum occurs at $(k_x = 0, k_y = 0)$, in agreement with the experimentally reported intralayer FM spin ordering[31]. For monolayers $NiBr_2$ and $NiI_2$ energy minima are located at $\boldsymbol{k} = (-0.06, 0.12.0)$ and $\boldsymbol{k} = (-0.11, 0.22.0)$ respectively, in agreement with the expected non-collinear helimagnetic spin ordering. It is noteworthy to mention that SOC effects, although relevant for the proper description of magnetic properties and determination of the ground-state of $NiI_2$, as recently reported in[34], are not crucial for our purposes.

In order to identify the primary coupling mechanisms leading to the detected spin states, we started by constructing a standard bilinear Heisenberg spin Hamiltonian:

$$H_J = \sum_{<i,j>} J_1 S_i \cdot S_j + \sum_{<i,l>'} J_2 S_i \cdot S_l + \sum_{<i,k>''} J_3 S_i \cdot S_k. \qquad (1)$$

where $J_1$, $J_2$ and $J_3$ are first-nearest-neighbor (NN), second-NN and third-NN Heisenberg exchange parameters, respectively. Using a 5×4×1 supercell of monolayer $NiX_2$, we computed these interaction parameters by means of the four-state method[18,39]. The results for $NiX_2$ monolayer, summarized in **Table 1** (negative and positive values represent FM and AFM interactions, respectively), show NN FM exchange interactions $J_1$ and third-NN AFM interactions $J_3$ to be strong, whereas the small second-NN FM interactions $J_2$ can be neglected.

We then employed the classical-spin analysis based on Fressier method[40] and parallel tempering Monte-Carlo (PTMC) simulations[41] based on the Heisenberg $J_1$~$J_2$~$J_3$ $H_J$ model, to check if such a model correctly describes the spin energetics, as resulting from the gBT. Incidentally, we note that, due to negligible $J_2$ in $NiX_2$, the role of the AFM $J_2$ interaction in the classical phase diagram for the Heisenberg $J_1$~$J_2$ model for the triangular lattice[42,43] is here played by a strong AFM $J_3$. As reported in **FIG.S4**, the plot of $E(k)$ vs $k$ curve for the monolayer $NiX_2$ based on the Fressier method, using the two dominant exchange interactions $J_1$ and $J_3$, shows that the minimum energy points occur at $k$ = (-0.13,0.26) for $NiI_2$ and $k$ = (-0.09,0.18) for $NiBr_2$, in agreement with the gBT results (**FIG.2**(b) and **FIG.S7** respectively). On the other hand, the $E(k)$ curve for monolayer $NiCl_2$ also shows minima at (-0.06, 0.12), indicating that the lowest energy spin configuration should also be a spin spiral. This is inconsistent with the above gBT results and with the experimentally observed FM ($k_x$ = $k_y$ = 0) arrangement of spins within a single-layer of $NiCl_2$[44]. To further check the results from the Fressier method, we performed PTMC simulations [see **FIG.S5**(a-c)] based on the minimal Heisenberg $J_1$~$J_2$~$J_3$ spin Hamiltonian, finding out consistent results: monolayer $NiCl_2$ would adopt the AFM non-collinear spiral order, in contradiction to the experimental and first-principles gBT results. This suggests that the bilinear Heisenberg model (i.e., $J_1$~$J_2$~$J_3$ model) fails to describe the stability of the FM spin state in monolayer $NiCl_2$ and that higher-order exchange coupling interactions have to be considered.

To extract other important exchange parameters, we employed our newly developed Machine Learning Method for Constructing Hamiltonian (MLMCH)[32] which can automatically select out the important ones among many possible interactions[45]. As seen from **Table.1**, the bilinear exchange interactions ($J_1$, $J_2$ and $J_3$) fitted by MLMCH are close to the values directly estimated via the four-state method, thus indicating the reliability of our approach. Most notably, our MLMCH process suggests that the NN biquadratic interaction $K$, reported in **Table. 1**, is significant in monolayer $NiCl_2$. The $K$ term is in fact of the same order of magnitude as $J_3$ in $NiCl_2$ with $|K/J_3|$~0.6, while the $|K/J_3|$ ratio significantly decreases to 0.3 and 0.2 in $NiBr_2$ and $NiI_2$, respectively. To further validate the K term estimate and its relevance as coupling parameter in $NiCl_2$, we calculate $J_1$ and $K$ by both an element substitution method and by means of the Liechtenstein, Katsnelsson, Antropov and Gubanov (LKAG) formalism [46], as detailed in the SM[33]. We found confirmation of the above predicted importance of the biquadratic interaction. Such results thus indicate that the biquadratic K contribution cannot be neglected when exploring magnetic properties in $NiCl_2$. We note that the importance of the biquadratic interaction is unraveled from our MLMCH process, in contrast to the recent study on some other layered magnetic materials where the form of the spin Hamiltonian was assumed from the very beginning[47].

To clarify the role of biquadratic interaction on the magnetic ground state of monolayer NiCl$_2$, we carried out PTMC simulations with $H_{JK}$ which was obtained after adding the $K$ term to the $J_1$~$J_2$~$J_3$ Heisenberg Hamiltonian as follows:

$$H_{JK} = \sum_{<i,j>}[J_1 S_i \cdot S_j + K(S_i \cdot S_j)^2] + \sum_{<i,l>'} J_2 S_i \cdot S_l + \sum_{<i,k>''} J_3 S_i \cdot S_k \quad (2)$$

Surprisingly and interestingly, the magnetization-temperature (M-T) curve presented in **FIG. S5**(a) of the SM[33] shows that the lowest-energy spin arrangement in NiCl$_2$ develops now a net magnetization M, thus corresponding to a FM order, whereas no such effects are observed in NiBr$_2$ and NiI$_2$ [**FIG.S5**(b,c)]. Therefore, the negative NN biquadratic interaction $K$ plays a key role in the stabilization of the FM state in monolayer NiCl$_2$, favoring collinear spin ordering and weakening the effect towards non-collinearity, as induced by the competing AFM $J_3$ and FM $J_1$ interactions. On the other hand, in NiBr$_2$ and NiI$_2$ the spin exchange frustration is much stronger than that in NiCl$_2$, so that the biquadratic interaction does not have any relevant effect on the ground state (see **FIG.S5**[33]). It is noteworthy to mention that we adopted the PTMC simulations to investigate the zero-temperature magnetic ground state of monolayer NiX$_2$ while the finite temperature effects on the magnetic properties are not considered in this work. In addition, we calculate relative energies for the FM and ***k*** = (-0.06, 0.12) spiral states for NiCl$_2$ to find that the FM state has a lower energy by 0.12 meV/Ni using the $H_{JK}$ model.

As demonstrated above, the NN biquadratic exchange interaction $K$ is strongly negative and plays an essential role in monolayer NiCl$_2$, but its origin is still unclear. Previous studies[48] based on the two-orbital Hubbard model report a positive biquadratic exchange interaction $K$ for $S = 1$ systems. To resolve this controversy, we employed a three-orbital Hubbard model to derive the NN bilinear exchange interaction $J$ and biquadratic exchange interaction $K$ through exact diagonalization. For simplicity, we consider a two-site cluster with four electrons, similar to Mila *et al.*[49]. As shown in the inset of **FIG.3**(a), each site provides three orbitals; the third empty atomic orbital has higher energy than the two singly occupied degenerate orbital states by an amount $\Delta$ (*e.g.*, due to crystal field splitting). The ground state of the cluster satisfies Hund's first rule, *i.e.*, each site adopts a high-spin configuration. The multi-orbital Hubbard model Hamiltonian can be written as:

$$H = \sum_{\alpha \neq \alpha',\sigma} t_{\alpha\alpha'}(c^+_{1,\alpha,\sigma} c_{2,\alpha',\sigma} + H.c.) + \sum_{i,\sigma} \Delta(c^+_{i,3,\sigma} c_{i,3,\sigma}) + H_D, \quad (3)$$

where the first term represents electron hopping from orbital $\alpha$ at site-1 to orbital $\alpha'$ at site-2 with spin conserved, the second term is the onsite energy of the third empty orbital relative to the two low energy degenerate orbitals, the last term is the Hubbard repulsion term (see SM[33]). For the two-site cluster, the low energy effective spin Hamiltonian $H_{ij} = J(S_i \cdot S_j) + K(S_i \cdot S_j)^2$ is determined by the total spin $S$ of two neighboring sites $i$ and $j$. $E_S$ denotes the energy of the two-site cluster with total spin $S$ ($S$ = 0,1 or 2). Then, the $J$ and $K$ parameters can be written in terms of $E_0$, $E_1$ and $E_2$ as: $J = (E_2 - E_1)/2$, and $K = E_0/3 + E_2/6 - E_1/2$ (see the SM[33] for detailed derivation). For simplicity, we adopt $t_{11}=t_{33}=t_{12}=t_{13}=0$, $t_{23}=xt_{22}$ (with parameter $x$ = 0~2). Here we set $t_{22}$ = 0.3 eV, $\Delta$ = 0.5 eV, $U$ = 5 eV, $J_H$ = $U$/5, $U$ = $U'$ + 2$J_H$, and we diagonalize the Hamiltonian. From **FIG.3**(a), we see that the NN biquadratic exchange interaction is close to zero when $t_{23}/t_{22}$ is small, whereas it becomes large when $t_{23}/t_{22}$ is greater than 1. Our exact diagonalization results are also confirmed by our perturbation analysis (not shown here). Note that the NN biquadratic exchange interaction is negative, at variance with the results of the two-orbitals Hubbard model[48]. The relative strength of the biquadratic interaction $\beta$ (-$K/J$) is close to zero, when $t_{23}/t_{22}$ is small; on the other hand, $\beta$ diverges for $t_{23}/t_{22}$ approaching 1. Furthermore, $\beta$ gradually converges to -0.5 when $t_{23}/t_{22}$ is greater than 1. The main conclusions on $J$ and $K$ remain qualitatively unaltered for different parameters (see **FIG.S9-11** in SM[33]).

In the case of the $Ni^{2+}$ ($d^8$) ion in $NiX_2$, there are two unoccupied $d$ orbitals. Within the hole picture, we reverse the energy levels in the crystal field, hence $e_g$ and $t_{2g}$ exchange their order. The two singly occupied orbitals are now $e_g$ orbitals, whereas the third empty orbital is one of the three $t_{2g}$ orbitals. The gap $\Delta$ is due to the $t_{2g}$-$e_g$ crystal field splitting. The hopping between two $Ni^{2+}$ ions is mediated by the ligand X ions, while the direct hopping between two $Ni^{2+}$ is usually small (as they are far apart) and can thus be neglected. In order to employ the aforementioned three-orbital Hubbard model, we designate $d_{x2-y2}$ as orbital-1, $d_{z2-3r2}$ as orbital-2 and $d_{xy}$ as orbital-3 (for simplicity we select $d_{xy}$ among the three $t_{2g}$ orbitals as the empty orbital). We estimate $t_{23}/t_{22}$ in the Ni-Cl-Ni cluster. The Ni-Cl-Ni bond angle in $NiCl_2$ with edge-shared octahedra is close to 90°. According to the interatomic Slater-Koster (SK) two-center integrals[50], we can identify the possible indirect hopping configurations between inter-sites $d$-shells. For the Ni-Cl-Ni cluster with 90° bond, a large hopping occurs between the $e_g$ orbitals and empty $d_{xy}$ orbital, while the hopping between the $e_g$ and $e_g$ orbitals, $d_{xy}$ and $d_{xy}$ orbitals almost vanishes (i.e., $t_{22}$ is close to 0, though not exactly 0 due to the direct hopping between the $d$ orbitals), suggesting that $t_{23}/t_{22}$ is gigantic (see SM[33] for details). Therefore, the giant NN biquadratic exchange interaction of monolayer $NiCl_2$ can be explained in terms of the large hopping ratio $t_{23}/t_{22}$.

Our analysis shows that there is a giant biquadratic exchange interaction in the Ni-Cl-Ni cluster with the 90° bond angle due to the large $t_{23}/t_{22}$, suggesting to investigate the dependence of the biquadratic exchange interaction on the bond angle between transition metal (TM) and ligand. Let us first consider the 180° bond angle case. In contrast to the 90° bond angle case, hopping only occurs between the $e_g$ and $e_g$ orbitals, which suggest $t_{23}/t_{22}$ to be zero, i.e. the NN biquadratic exchange interaction is close to zero in the Ni-Cl-Ni cluster with the 180° bond angle. This also explains why the biquadratic exchange interaction is usually small and/or not important in perovskites, where the TM-ligand-TM angle is close to 180°. For further insights, we also plot $t_{23}/t_{22}$ as a function of the Ni-Cl-Ni bond angle [see **FIG.3**(b)]. As evident, $t_{23}/t_{22}$ changes slowly, when the bond angle approaches 180°, whereas $t_{23}/t_{22}$ increases sharply, when the bond angle is close to 90°. To verify the tight binding (TB) analysis, we calculated by means of DFT the relative biquadratic exchange interactions $-K/J$ as a function of the Ni-Cl-Ni bond angle, in a fictitious perovskite $NaNiCl_3$ with various octahedral rotations. The DFT results show $-K/J$ to be close to zero when the Ni-Cl-Ni bond is close to 180°, while $-K/J$ is larger than 0.5 when the bond angle approaches 90°, in line with our theoretical expectations based on the TB analysis.

Besides $NiX_2$ with a triangular lattice, TM-ligand-TM arrangements with an approximate 90° bond angle is common in 2D magnets with honeycomb lattice, such as $CrI_3$ and $CrGeTe_3$. The latter systems have attracted numerous attention [6,7], but whether the biquadratic exchange interaction is important is still unclear. We thus calculated NN bilinear and biquadratic exchange interactions $K$ in monolayer $CrCl_3$, $CrI_3$, $CrSiTe_3$ and $CrGeSe_3$, finding out large values. In addition, we also found that $K$ is not sensitive to the applied strain, at variance with the NN $J$. Indeed, as shown in **FIG.S15**, $J$ changes sharply when the applied strain changes from -4% to 4%, whereas $K$ is almost unaffected, generally indicating $K$ to be important in magnets with edge-shared octahedra. Compared to $J$, the biquadratic exchange $K$ is less sensitive to the hopping parameters, which explains why $K$ is not very sensitive to strain.

***Discussion and Summary***-Previous experiments in NiO [51] show that it is necessary to take the NN biquadratic interaction $K$ into account to explain the observed anomalous behavior of the sublattice magnetization. However, the experimentally fitted $|K/J_1|$ is about 0.05, i.e. much smaller than that of monolayer $NiCl_2$. For monolayer $NiCl_2$, our calculations show that giant negative NN biquadratic interaction overcomes the spin frustration caused by nearest-neighbor FM and third-neighbor AFM exchange interactions, reproducing the experimentally observed intralayer FM spin ordering. As such, it represents an unconventional situation, where the FM spin order is stabilized by the competition between giant nearest-neighbor biquadratic and bilinear exchange interactions. Finally, our TB analysis shows that the 90° TM-ligand-TM motif plays a key role, suggesting that the biquadratic exchange interactions may be relevant also in other magnetic systems with edge-shared octahedra.

**Acknowledgments.** We thank Dr. C. Xu and Dr. P. Barone for interesting discussions. Work at Fudan is supported by NSFC (11825403, 11991061), Program for Professor of Special Appointment (Eastern Scholar), and Qing Nian Ba Jian Program. J.F. acknowledges the support from Anhui Provincial Natural Science Foundation (1908085MA10). Work at CNR-SPIN supported by Nanoscience Foundries and Fine Analysis (NFFA-MIUR Italy) and by the PRIN-2017 project "TWEET: Towards Ferroelectricity in two dimensions" (IT-MIUR Grant No. 2017YCTB59). XH thanks the support by the EU H2020-NMBP-TO-IND-2018 project "INTERSECT" (Grant No. 814487).

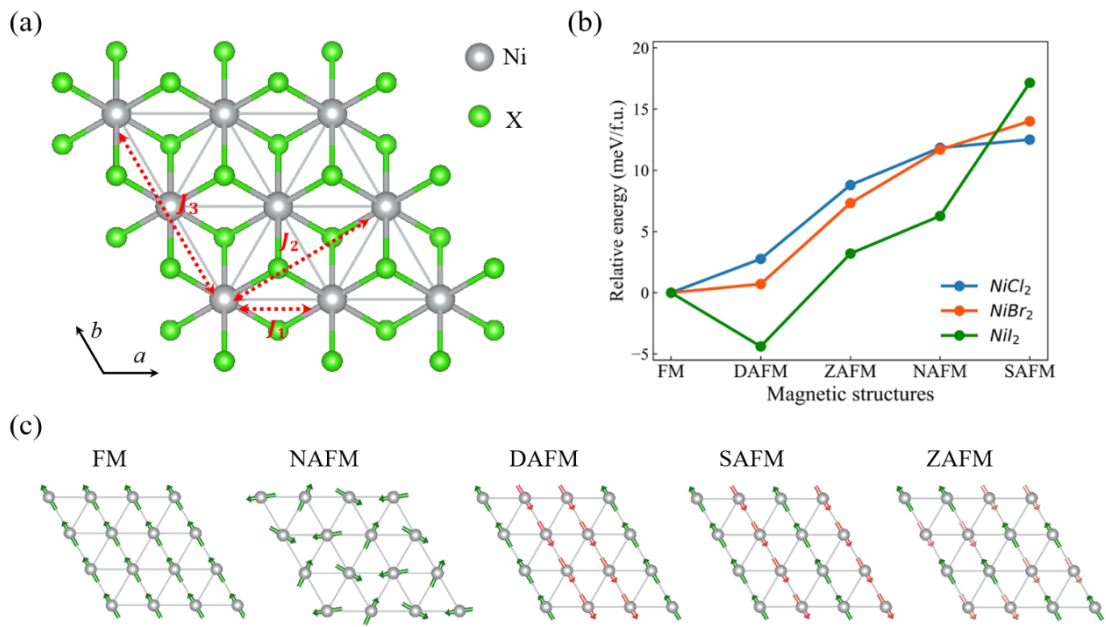

**FIG. 1** (a) Atomic arrangement of monolayer NiX$_2$, along with magnetic exchange paths (i.e. first-neighbor ($J_1$), second-neighbor ($J_2$) and third-neighbor ($J_3$)). (b) Calculated relative energies for various magnetic structures of monolayer NiX$_2$ by GGA+U method using the structure optimized with the FM order. Here, the FM state is chosen as energy reference. (c) Schematic top view of various magnetic structures.

**Table. 1** The bilinear exchange interactions (in meV) and NN biquadratic interaction (in meV) calculated with MLMCH (or four-state method) for monolayer $NiX_2$. The SOC effect is not included here.

| Pairs | $J_1$ | $J_2$ | $J_3$ | $K$ |
| --- | --- | --- | --- | --- |
| $NiCl_2$ | -3.17(-3.08) | -0.04(-0.04) | 0.88(0.91) | -0.54 |
| $NiBr_2$ | -3.75(-3.46) | -0.09(-0.08) | 1.59(1.63) | -0.51 |
| $NiI_2$ | -4.54(-4.23) | -0.19(-0.17) | 3.37(3.38) | -0.59 |

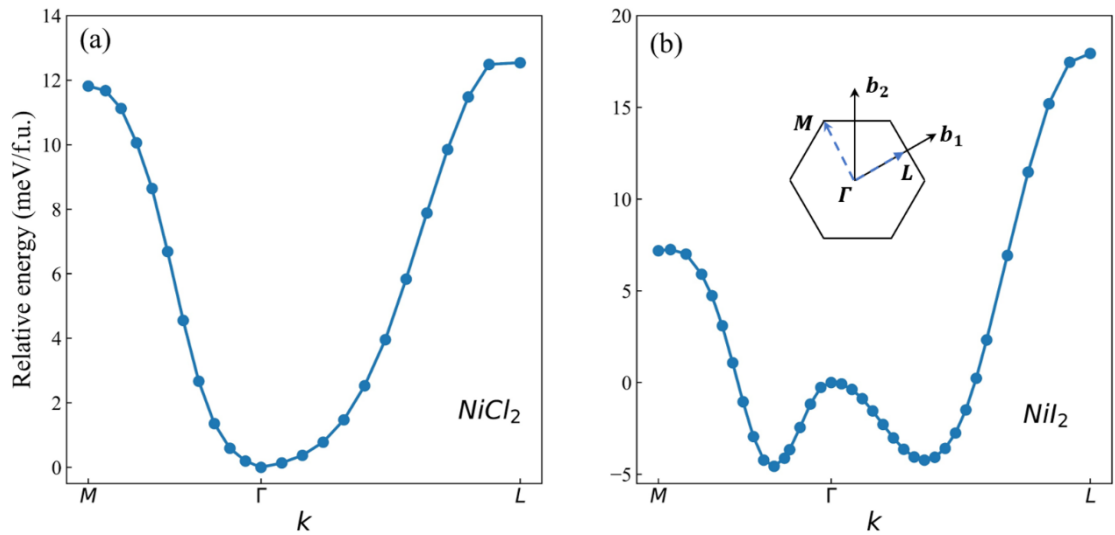

**FIG. 2** (a)-(b) The relative total-energy dependence on the magnetic propagation vector $k$ for monolayer NiCl$_2$ and NiI$_2$, respectively. The FM state (*i.e.*, $k = \Gamma$) is chosen as the reference.

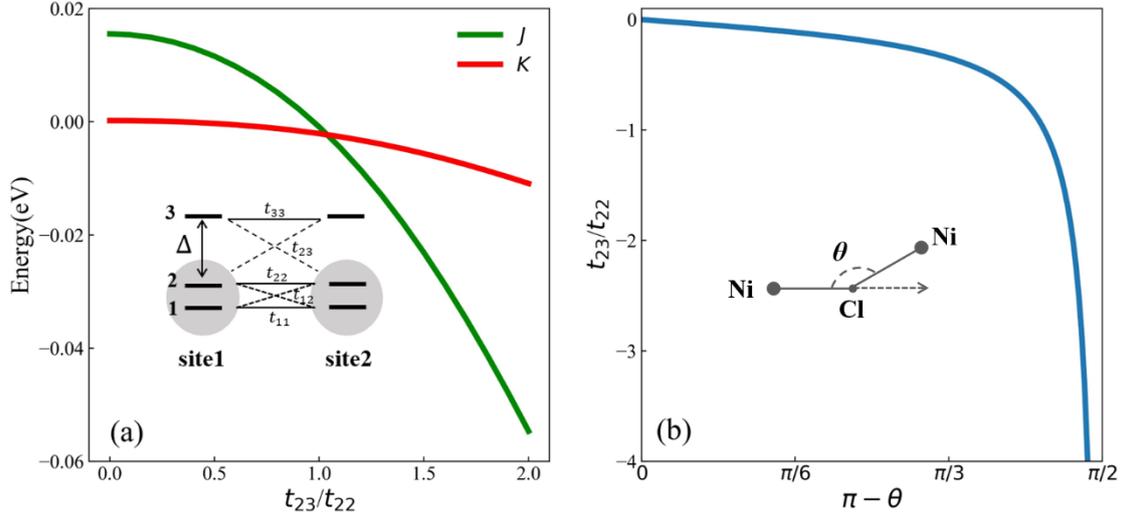

**FIG. 3** (a) $J$ (NN bilinear exchange interaction) and $K$ (NN biquadratic exchange interaction) from the TB calculations as a function of $t_{23}/t_{22}$. (b) Relative amplitude $t_{23}/t_{22}$ as a function of $\pi$-$\theta$ ($\theta$ denotes the Ni-Cl-Ni bond angle).

given cutoff distance (11 Å in this work).